# Controlling the Emission of Electromagnetic Sources by Coordinate transformation

Yu Luo [1], Jingjing Zhang [1], Lixin Ran [1]*, Hongsheng Chen [1,2], and Jin Au Kong [1,2]

[1] *The Electromagnetics Academy at Zhejiang University, Zhejiang University, Hangzhou 310058, P. R. China*

[2] *Research Laboratory of Electronics, Massachusetts Institute of Technology, Cambridge, Massachusetts 02139*

**Abstract**

The coordinate transformation on the space that contains electromagnetic sources is studied. We find that, not only the permittivity and permeability tensors of the media, but also the sources inside the media will take another form in order to behave equivalently as the original case. It is demonstrated that, a source of arbitrary shape and position in the free space can be replaced by an appropriately designed metamaterial coating with current distributed on the inner surface and would not be detected by outer observers, because the emission of the source can be controlled at will in this way. As examples, we show how to design conformal antennas by covering the sources with transformation media. The method proposed in this letter provides a completely new approach to develop novel active EM devices.

*\* Author to whom correspondence should be addressed; electronic mail: ranlx@zju.edu.cn*

In 2006, J. Pendry et al. proposed the pioneering coordinate transformation concept to arbitrarily control electromagnetic (EM) fields [1, 2]. In his approach, a space consisting of the normal free space can be transformed into a new space with different volume and space-distributed constitutive parameters [1, 3, 4], which could be realized by artificial metamaterial, whose permittivity and/or permeability can be designed to continuously change from negative to positive values[5-7]. Following this approach, some novel EM devices — one example is the famous invisibility cloak — can be realized to obtain unusual EM behaviors [8-20]. In this Letter, we further illustrate that such coordinate transformation can even be applied to a space containing EM sources. The closed-form solution obtained from a full wave analysis that combines the coordinate transformation and green's function principle shows that in this case, not only the permittivity and permeability tensors, but also the EM sources can be transformed at will. In particular, a source of arbitrary shape and position in the free space can be replaced by another source, given access to a properly designed metamaterial coating, without being detected by outer observers. As examples to illustrate the possible applications of this kind of transformation, we designed two conformal antennas — which always attracts interests of microwave engineers — by transforming a simple dipole antenna and a standard array antenna, separately. The closed form formulations and the calculations show that the metamaterial coatings carrying specific currents have the same radiation patterns as the original antennas' in the zones of interest. The method proposed in this Letter provides a brand-new approach in the development of novel active EM devices, and enlightens a wider exploration in the field of the EM coordinate transformation.

We start from considering a general coordinate transformation between the initial and the transformed coordinate systems ($x', y', z'$) and ($x, y, z$) by

$$\vec{r}' = \vec{F}(\vec{r}) \quad \text{or} \quad x' = f(x,y,z) \quad y' = g(x,y,z) \quad z' = h(x,y,z) \ . \tag{1}$$

Here $\vec{r}' = x'\hat{x} + y'\hat{y} + z'\hat{z}$, $\vec{r} = x\hat{x} + y\hat{y} + z\hat{z}$, and $\overline{F}$ represents a deformation mapping from coordinate system $(x', y', z')$ to $(x, y, z)$ on a closed domain $V$, of which the boundary is matched to free space (i.e., $\vec{r} = \vec{F}(\vec{r})\big|_{\partial V}$). The relation between the two operators $\tilde{\nabla} = \hat{x}\frac{\partial}{\partial x'} + \hat{y}\frac{\partial}{\partial y'} + \hat{z}\frac{\partial}{\partial z'}$ and $\nabla = \hat{x}\frac{\partial}{\partial x} + \hat{y}\frac{\partial}{\partial y} + \hat{z}\frac{\partial}{\partial z}$ is given by

$$\begin{cases} \nabla \times \overline{A} = \det(\overline{\overline{J}}) \overline{\overline{J}}^{-1} \cdot \tilde{\nabla} \times \left[ \left(\overline{\overline{J}}^{\mathrm{T}}\right)^{-1} \cdot \overline{A} \right] \\ \nabla \cdot \overline{A} = \det(\overline{\overline{J}}) \tilde{\nabla} \cdot \left[ \overline{\overline{J}} \cdot \overline{A} \big/ \det(\overline{\overline{J}}) \right] \end{cases}, \tag{2}$$

where $\overline{\overline{J}} = \frac{\partial(f,g,h)}{\partial(r,\theta,\varphi)}$ is the Jacobian transformation matrix [3][4]. Substituting equations (2) into Maxwell's equations, we obtain the desired EM constitutive parameters in terms of the constants in the original coordinate

$$\begin{cases} \overline{\overline{\varepsilon}} = \det(\overline{\overline{J}}) \overline{\overline{J}}^{-1} \cdot \overline{\overline{\varepsilon}}' \cdot \left(\overline{\overline{J}}^{\mathrm{T}}\right)^{-1}, \quad \overline{\overline{\mu}} = \det(\overline{\overline{J}}) \overline{\overline{J}}^{-1} \cdot \overline{\overline{\mu}}' \cdot \left(\overline{\overline{J}}^{\mathrm{T}}\right)^{-1} \\ \rho(\vec{r}) = \det(\overline{\overline{J}}) \rho'(\overline{F}(\vec{r})), \quad \overline{j}(\vec{r}) = \det(\overline{\overline{J}}) \overline{\overline{J}}^{-1} \cdot \overline{j}'(\overline{F}(\vec{r})) \end{cases}. \tag{3}$$

The surface current $\overline{J}_s$, current $\overline{I}$, and charge $q$ in the transformed coordinate can also be described with the mapping matrix $\overline{F}$ as

$$\overline{J}_s(\vec{r}) = \overline{\overline{J}}^{\mathrm{T}} \cdot \overline{J}_s'(\overline{F}(\vec{r})), \quad \overline{I}(\vec{r}) = \overline{I}'(\overline{F}(\vec{r})), \quad q(\vec{r}) = q'(\overline{F}(\vec{r})). \tag{4}$$

Note that Maxwell's equations still remain their form invariance in the new coordinates $(x, y, z)$, but the permittivity, permeability, charge density and current density will be with different forms, and the total charge $q$ and the total current $\overline{I}$ satisfy conservation law under the transformation. To evaluate the electromagnetic field in the transformed coordinates on domain $V$ which contains sources $\overline{j}(\vec{r}_0)$ and

$\rho(\bar{r}_0)$, we introduce the vector potential $\bar{A}$ ($\bar{B} = \nabla \times \bar{A}$) and scatter potential $\phi$ ($\bar{E} = i\omega\bar{A} - \nabla\phi$). By applying gauge condition $\nabla \cdot (\bar{\bar{T}}^{-1} \cdot \bar{A}) - i\omega\mu_0\varepsilon_0\phi = 0$, where $\bar{\bar{T}} = \bar{\bar{J}}^T \bar{\bar{J}} / \det(\bar{\bar{J}})$ is the metric tensor, the partial differential equation of the vector potential $\bar{A}$ can be obtained, and thereby the exact expression of the electric field can be written with Green's function

$$\bar{E} = i\omega\mu_0 \bar{\bar{J}}^T \cdot \left(\bar{\bar{I}} + \frac{\tilde{\nabla}\tilde{\nabla}}{k^2}\right) \cdot \iiint d^3\bar{r}_0' \frac{e^{ik|\bar{r}'-\bar{r}_0'|}}{4\pi|\bar{r}'-\bar{r}_0'|\det(\bar{\bar{J}})} \bar{\bar{J}} \cdot \bar{j}(\bar{r}_0), \tag{5}$$

with $\bar{r}' = \bar{F}(\bar{r})$, $\bar{r}_0' = \bar{F}(\bar{r}_0)$.

The above analysis implies that, by applying the coordinate transformation $\bar{F}$, the point $\bar{r}' = \bar{F}(\bar{r}_0)$ in the initial coordinates will be transformed to $\bar{r} = \bar{r}_0$ in the new coordinates. Thus looking from the outside of the transformed region, a source located at $\bar{r} = \bar{r}_0$ seems to radiate at $\bar{r} = \bar{F}(\bar{r}_0)$. Then theoretically, by choosing appropriate transformation $\bar{F}$, a given source can be transformed into any required shape and position. This principle can be applied in the design of active EM devices, especially antennas, which will be raised as examples in the following part of this Letter.

J. Pendry et al. suggested that with certain transformation, the electromagnetic field in a spherical region $r < R_2$ can be compressed into an annulus region $R_1 < r < R_2$ to form an invisibility cloak [1]. Here we consider a similar situation but the mapping $\bar{F}$ is selected by

$$\frac{x'^2 + y'^2}{a^2(r)} + \frac{z'^2}{b^2(r)} = 1, \tag{6}$$

where $a(r) = \frac{R_2}{R_2 - R_1}(r - R_1)$ and $b(r) = \frac{R_2 - \frac{d}{2}}{R_2 - R_1}(r - R_1) + \frac{d}{2}$ (Here $d$ is a constant).

Under this transformation, the outer boundary of the annulus ($r = R_2$) still matches the free space, which is similar to the cloaking case, however, the thing different is that a line with a length $d$ with respect to the original ($x', y', z'$) axes, instead of the origin, is transformed to the spherical inner surface $r = R_1$. Suppose there is a dipole with length $d$ located at the origin, as shown in Fig. 1(a). After the proposed transformation, the line current carried by the dipole will be mapped to the surface current on the inner boundary ($r = R_1$), and in the spherical region $r < R_2$, the electric field radiated by the dipole will be squeezed into the annulus $R_1 < r < R_2$. We can comprehend this from another perspective: suppose there are some currents distributed along the $\theta$ direction on the inner surface $r = R_1$, and the region $R_1 < r < R_2$ is filled with this transformed medium. Looking from the outer space $r > R_2$, a detector will only 'see' a dipole of length $d$ pointing in the $z$ direction and the radiation field is completely the same as that of the dipole. This also provides a simple way to understand Pendry's cloaking [1]. In Pendry's transformation, a point in the original space is transformed to the inner spherical surface in the new space, therefore, the currents distributed on the inner spherical surface $r = R_1$ simply behave as a point with neither net charges nor net dipole moment to outer observers. That's why under Pendry's transformation the current on the inner surface cannot radiate power and is also invisible for detections. Fig. 1(b) displays the transformed surface current $\bar{J}_s = -\frac{I_0}{2\pi R_1 \sin\theta}\hat{\theta}$ uniformly distributed along $\varphi$ direction on a spherical surface which works equivalently to a dipole under the aforementioned transformation (In fact, as long as the total current $\int_0^{2\pi} 2\pi R_1 \sin\theta \cdot \bar{J}_s(\varphi) d\varphi$ is equal to $I_0$, the electric far field pattern will not be affected no matter how the current is distributed on the surface). It seems that there are no $\varphi$ component surface currents in the transformed domain. It is easy to understand that at $r = R_1$ only the current in the $\theta$ direction can produce radiation while current in the $\varphi$ direction will make no contribution to the radiation, because looking from the outside of the annulus, a current in the $\varphi$ direction will only act like one point source on the line.

The electric fields in the transformed space can be obtained from Equations (5). To compare above two cases, the radiations of the original dipole and the transformed antenna consisting of the metamaterial coating carrying the spherical surface current, as well as a subtraction between them are shown in fig. 2(a), 2(b) and 2(c), respectively, and we can notice that the field distribution outside the spherical region ($r > R_2$) are exactly the same. Therefore a conformal antenna with the same radiation pattern as a dipole but with a spherical shape is obtained.

We next consider how to map a circular plane of diameter $d$ to a spherical surface, which would yield an equivalent conformal array antenna. Suppose this finite plane is in the y-z plane and the transformation take a similar form

$$\frac{x'^2 + y'^2}{a^2(r)} + \frac{z'^2}{b^2(r)} = 1, \tag{7}$$

where $a(r) = \frac{R_2 - \frac{d}{2}}{R_2 - R_1}(r - R_1) + \frac{d}{2}$ and $b(r) = \frac{R_2}{R_2 - R_1}(r - R_1)$. This transformation maps the circular plane with respect to the original $(x', y', z')$ axes to a spherical surface $r = R_1$. Suppose there is a dipole array of N elements pointing in the y direction and placed along the $x$ axis in the original space (within the plane of diameter $d$). Under this transformation, the currents of the dipole array will be mapped to the surface current on $r = R_1$. In other words, detected from outer domain $r > R_2$, it will appear that the emission is produced by this dipole array. Fig. 3(a) is the schematic of the transformation given in Equation (7), where N is assumed to be 3. The currents distributed on $r = R_1$ which will behave as three dipoles (depicted in Fig.3 (a) by red crosses) when the region $R_1 < r < R_2$ is composed of the proposed transformed metamaterial can be calculated with Equation (4) and are shown in Fig. 3 (b).

With Equation (5), we also calculate the electric field distribution of three dipoles in the original space and the equivalent surface current in the transformed space, which are shown in Fig. 4 (a) and (b), respectively, and the subtraction between them is again plotted in Fig. 4 (c).

Design of conformal antennas is always an approach of interest to microwave engineers, creating antennas that conform to a surface whose shape is determined by considerations other than electromagnetic; for example, aerodynamic or hydrodynamic. The above two specific cases indicate the basic principle for the design of conformal antennas using coordinate transformation. Different from the traditional method based on Huygens' principle which requires to simultaneously realize both the electric and magnetic currents on the aperture, here we simply use a metamaterial radome to control the current, and consequently the radiation without introducing non-practical perfect magnetic conductor or magnetic current. Not restricted to the proposed spherical cases, the virtual sources can be replaced by the currents on an arbitrary surface as long as we find an appropriate mapping $\bar{F}$ ($f, g, h$). The outer surface of the radome is matched to free space while a source of arbitrary shape and at any position inside the radome can be mapped to the current distributed on the inner surface. As a result, the conformal antenna will have exactly the same radiation pattern as the original virtual antenna emitting in the free space.

In summary, coordinate transformation is not restricted to source-free cases. The transformation on domains containing EM sources can also be studied by manipulating the Maxwell's equations. Under transformations, not only the space but also the sources can be transformed, and given access to proper metamaterials, the initial source of almost any shape and location can be replaced by another source with the same radiation properties. This makes the realization of some active EM devices, such as conformal antennas illustrated in this Letter, a practical possibility.

This work is sponsored by the Chinese National Science Foundation under Grant Nos. 60531020 and 60671003, the NCET-07-0750, the China Postdoctoral Science Foundation under Grant No. 20060390331, the ONR under Contract No. N00014-01-1-0713, and the Department of the Air Force under Air Force Contract No.F19628-00-C-0002.

**Figure Captions**

FIG 1: (color online) (a) Schematic of the spatial coordinate transformation which maps a line source to the currents distributed on a spherical surface. Red dashed line is a dipole pointing in the z direction and placed at the origin with length d in the original space. Under the transformation given in Equation (6), a line of length $d$ (red dashed line) is transformed to the inner boundary $r = R_1$ (denoted by red solid curve). (b) Current distribution on a spherical surface in the transformed physical space which has the same radiation pattern to that of the dipole shown in subfigure (a).

FIG 2: (color online) (a) Field distribution of a dipole in free space. (b) Field distribution of the equivalent surface current which is distributed at the inner boundary of a spherical shell constructed by transformation medium. (c) The difference between the fields in (a) and (b).

FIG 3: (color online) (a) Schematic of the spatial coordinate transformation which maps a dipole array to the currents distributed on a spherical surface. Red dashed line is a dipole array of three elements pointing in the $y$ direction and placed along the $x$ axis with equal spacing in the original space. Under the transformation given in Equation (7), the inner boundary $r = R_1$ (denoted by red solid curve) is transformed to a circular plane of diameter $d$ (The dipole array is within this plane) in the x-y plane. (b) Current distribution on a spherical surface in the transformed physical space which has the same radiation pattern to that of the three dipoles shown in subfigure (a).

FIG 4: (color online) (a) Field distribution of a current sheet in free space. (b) Field distribution of the equivalent surface current which is distributed at the inner

boundary of a spherical shell constructed by transformation medium. (c) The difference between the fields in (a) and (b).

FIG. 1

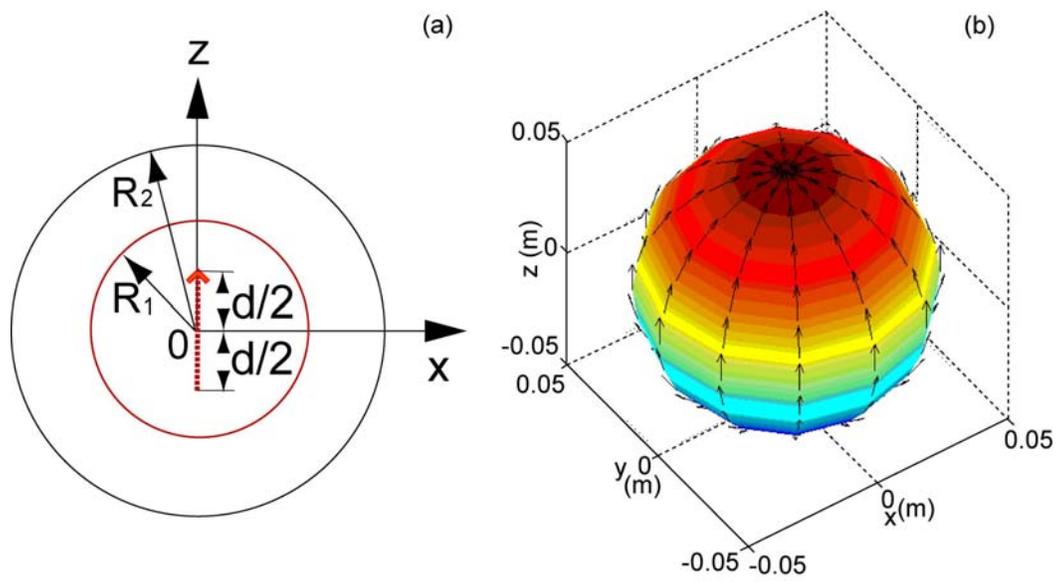

FIG. 2

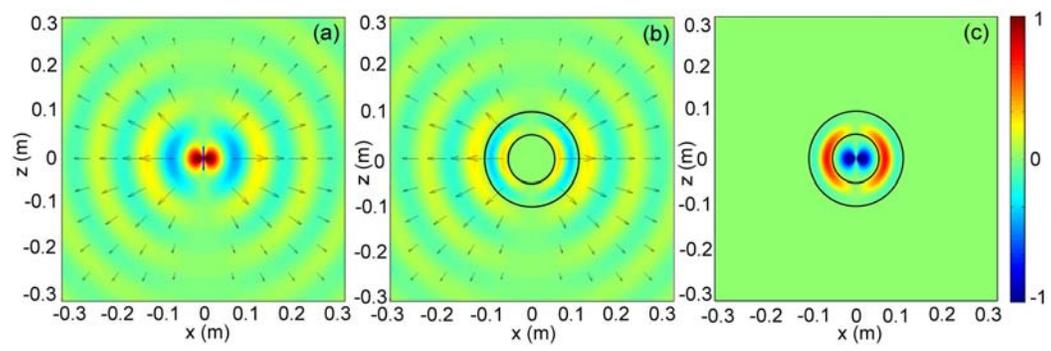

FIG. 3

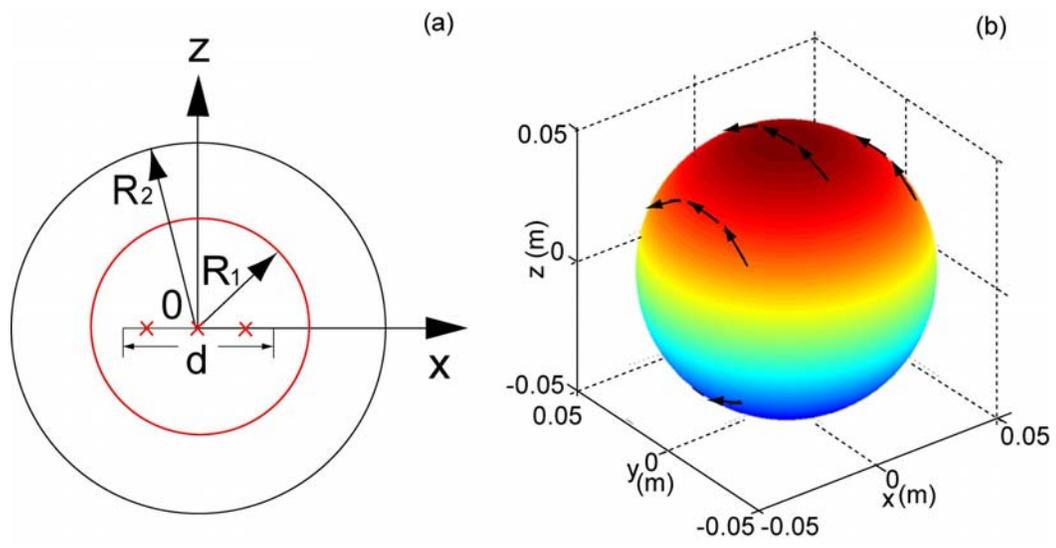

FIG. 4

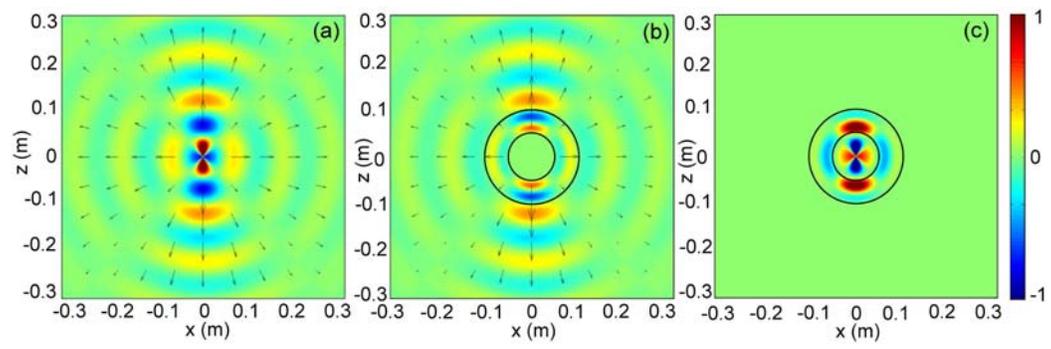